\documentclass[a4paper,11pt]{article}
\usepackage{pos}

\title{Energy Reconstruction and Calibration of the MicroBooNE LArTPC}

\author*[a, \dag]{Wanwei Wu}

\affiliation[a]{Fermi National Accelerator Laboratory,\\
  Batavia, Illinois 60510, USA}

\notes{\note{On behalf of the MicroBooNE Collaboration}}

\emailAdd{wwu@fnal.gov}

\abstract{The Liquid Argon Time Projection Chamber (LArTPC) is increasingly becoming the chosen technology for current and future precision neutrino oscillation experiments due to its superior capability in particle tracking and energy calorimetry. In LArTPCs, calorimetric information is critical for particle identification, which is the foundation for the neutrino cross-section and oscillation measurements as well as searches for beyond standard model physics. One of the primary challenges in employing LArTPC technology is characterizing its performance and quantifying the associated systematic uncertainties. MicroBooNE, the longest operating LArTPC to date, has performed numerous such measurements, including studies of detector physics and electromagnetic shower reconstruction. Here we present results on the operation and performance of the detector during its data taking, highlighting accomplishments toward calorimetric reconstruction, calibration, and detector physics.}

\FullConference{%
  *** The 23rd International Workshop on Neutrinos from Accelerators (NuFact 2022) ***\\
  *** 31 July - 6 August 2022 ***\\
  *** Salt Lake City, Utah, United States ***
}

\begin{document}

\maketitle

\section{Introduction}

MicroBooNE is a part of the Short-Baseline Neutrino (SBN) program at Fermilab \cite{sbn_fermilab}. The detector is an 85-tonne active volume liquid argon time projection chamber (LArTPC) located on-axis in the Booster Neutrino Beam (BNB) line. It sits $\sim 470$ m away from the target station and expects an on-axis flux of neutrinos with energies of approximately 0.2 GeV to 2 GeV \cite{BNB_flux}. It is also exposed to an off-axis flux of neutrinos from the NuMI neutrino beam \cite{NuMI}. MicroBooNE began collecting data in 2015 and ended its physics run in 2021. It acquired the largest statistics samples of neutrino interactions on argon to date.

The MicroBooNE LArTPC has an active volume of $2.56 \times 2.32 \times 10.36~\text{m}^{3}$ [drift coordinate ($x$), vertical ($y$), beam direction ($z$)]. Charged particles traversing the TPC can ionize the argon leaving trails of ionization electrons and produce scintillation light. The ionization electrons drift in an applied electric field (273 V/cm) to the three anode wire planes that provide the charge readout. The wire angles are $\pm 60^{\circ}$ from vertical for the two induction planes $U$ and $V$ and vertical for the collection plane $Y$ ($Y$ plane is the farthest from the cathode). The $U$, $V$, and $Y$ planes contain 2400, 2400, and 3456 wires, respectively. The plane spacing and wire pitch are both 3 mm. The $U$ and $V$ plane wires detect signal via induction, while the $Y$ plane wires collect the charge, which generally has the best resolution when converting charge to energy \cite{MicroBooNE_design}.

\section{Signal Processing}

\begin{figure}[ht!]
\begin{center}
\includegraphics[width=0.7\textwidth, trim = 0cm 0.1cm 0.0cm 0.2cm, clip]{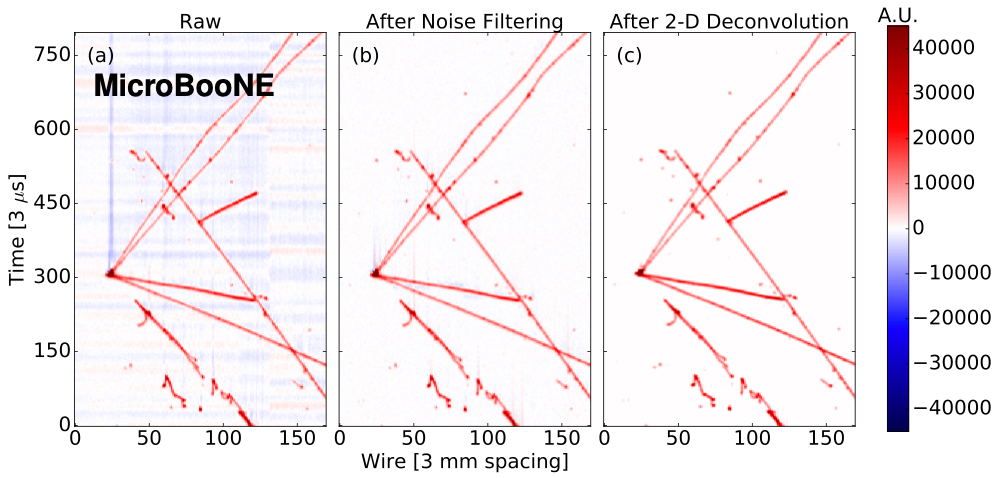}
\caption{Event displays of an example neutrino candidate from MicroBooNE data (run 3493, event 41075) showing a $Y$ plane view through each stage of signal processing. Figure is taken from Ref. \cite{SignalProcessing_microboone}}
\label{fig:signal_processing}
\end{center}
\vspace*{-0.1in}
\end{figure}

The charge information collected is recorded in the waveform from each wire channel. In the MicroBooNE LArTPC, common noise contains the coherent noise, tails, etc., mainly inherent to the electronics. The noise characteristics were studied in the frequency and time domains, and offline noise filters were developed to mitigate the noise \cite{noise_microboone}. The measured signal is a convolution of the original signal and detector response. For signal processing, a deconvolution method is used to extract the original signal from the measured signal to remove the detector response \cite{microboone_signalprocessing, SignalProcessing_microboone}. Figure \ref{fig:signal_processing} shows the event displays of an example neutrino candidate from MicroBooNE data at different stages: raw data, after noise filtering, and after deconvolution \cite{SignalProcessing_microboone}. The output of this deconvolution is a standard signal shape, i.e., a Gaussian shape, for the hit (charge) reconstruction.

\section{LArTPC Calibration and Energy Reconstruction}

\begin{figure}[ht!]
\begin{center}
\includegraphics[width=0.36\textwidth, trim = 0cm 0.1cm 0.0cm 0.1cm, clip]{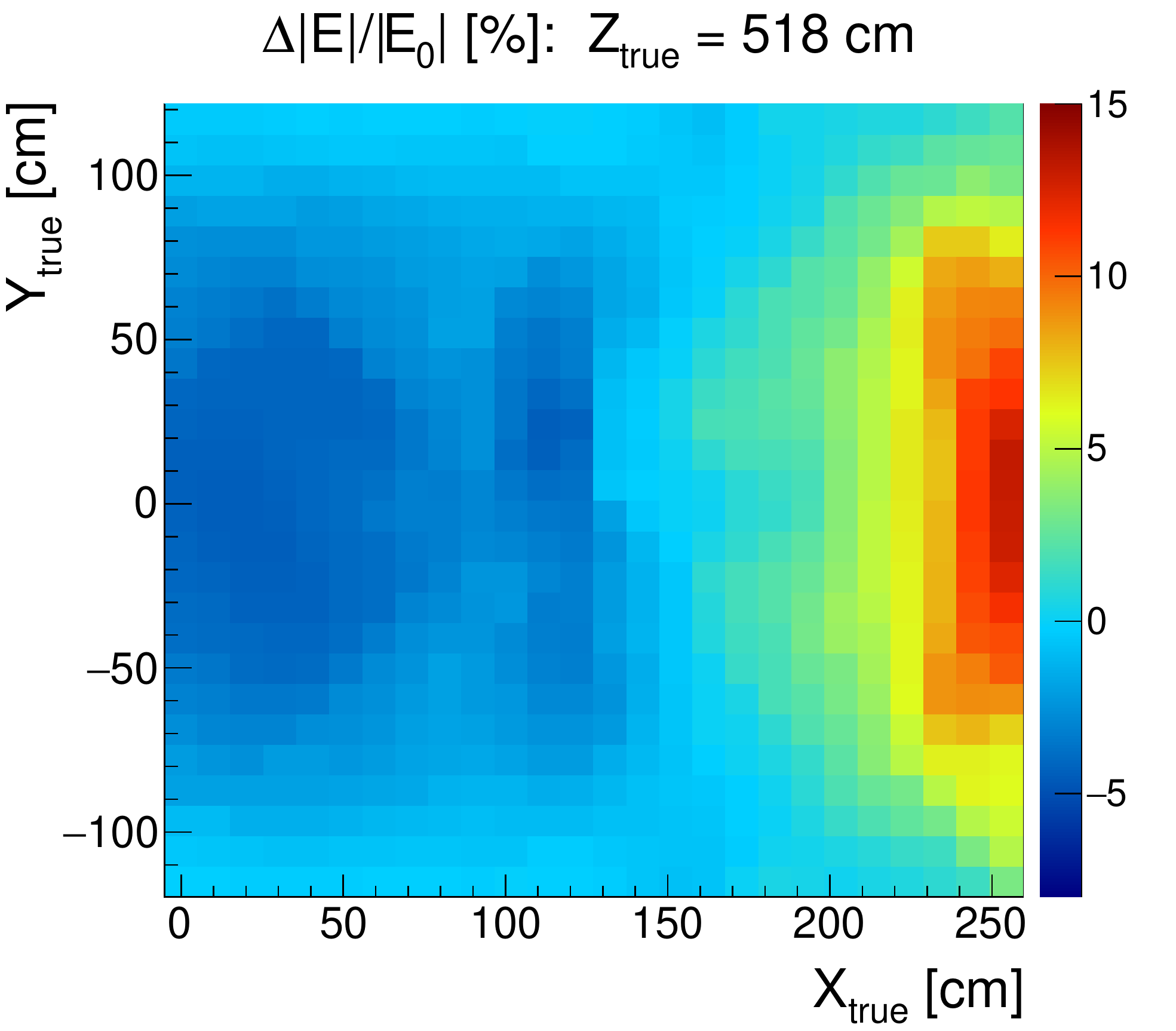}
\hspace*{0.3in}
\includegraphics[width=0.36\textwidth, trim = 0cm 0.1cm 0.0cm 0.1cm, clip]{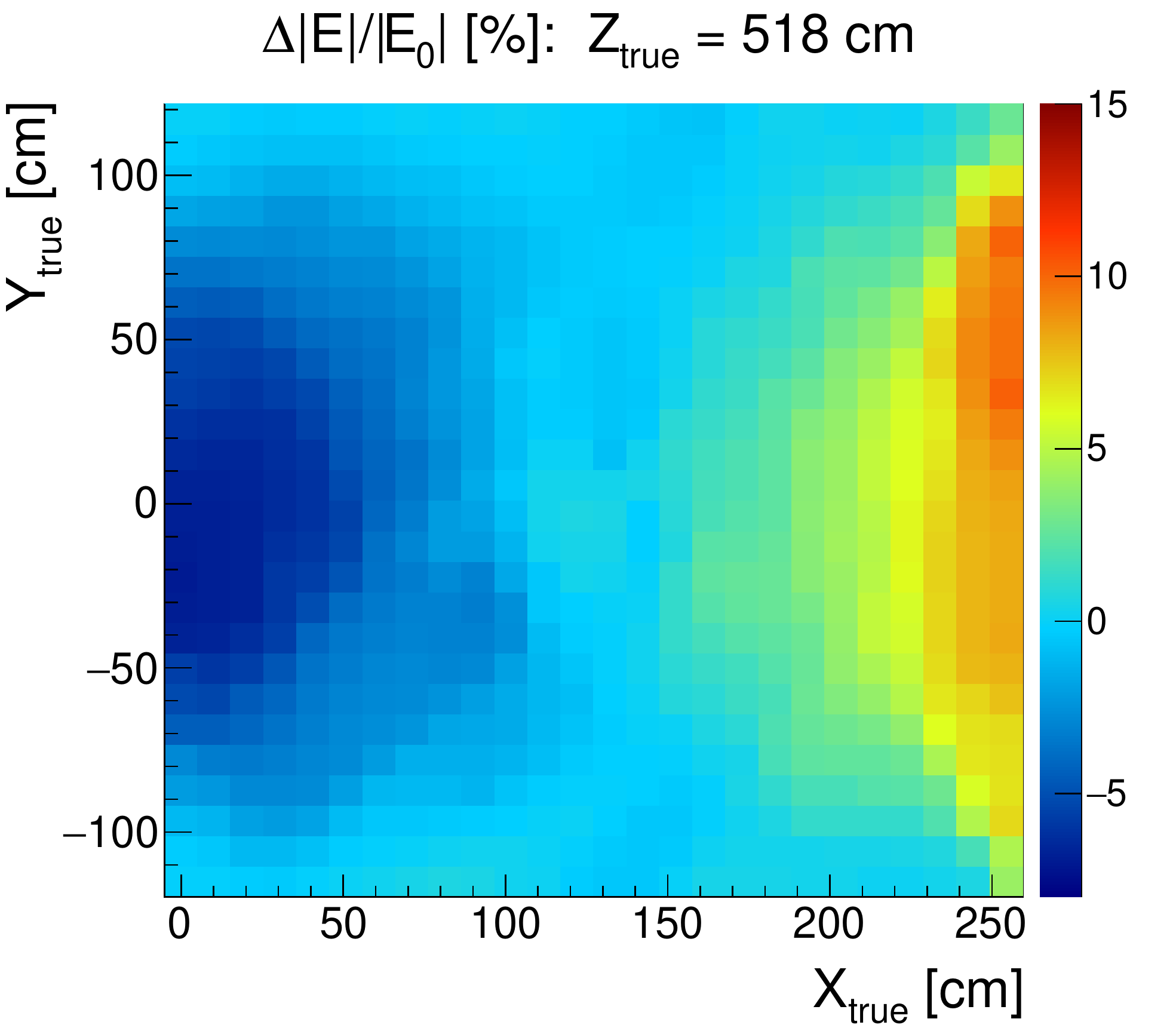}
\caption{Results of the calculation of electric field distortion magnitude for a central slice in $z$, comparing Monte Carlo simulation (left) to data (right). The distorn is due to SCE. Figure is taken from Ref. \cite{sce_microboone}.}
\label{fig:sce}
\end{center}
\end{figure}

LArTPC technology can provide excellent spatial and calorimetric resolutions. However, the total charge extracted after signal processing normally does not equal to the total charge produced from ionization. Distortions in detector response due to cross-connected TPC channels, space charge effects (SCE), electron attachment to impurities, diffusion, and recombination need to be corrected to get the exact amount of charge released from the original interaction \cite{calibration_microboone}. MicroBooNE is a near-surface detector. There are about $20-30$ cosmic muons per 4.8 ms readout window. The accumulation of slow-moving argon ions produced from cosmic muons impinging the TPC can distort the electrical field and the drift trajectories of ionization electrons. Such effects are known as SCE \cite{calibration_microboone, laser_microboone, sce_microboone}. Calibration of the electric field can be achieved by using the cosmic muon tracks \cite{sce_microboone} or the UV laser system \cite{laser_microboone}.
Figure \ref{fig:sce} shows an example electric field distortion magnitude for simulation and data using selected cosmic muon tracks \cite{sce_microboone}. The SCE has larger effects near cathode ($x=\sim256$ cm).

\begin{figure}
\begin{center}
\includegraphics[width=0.4\textwidth, height=0.27\textwidth]{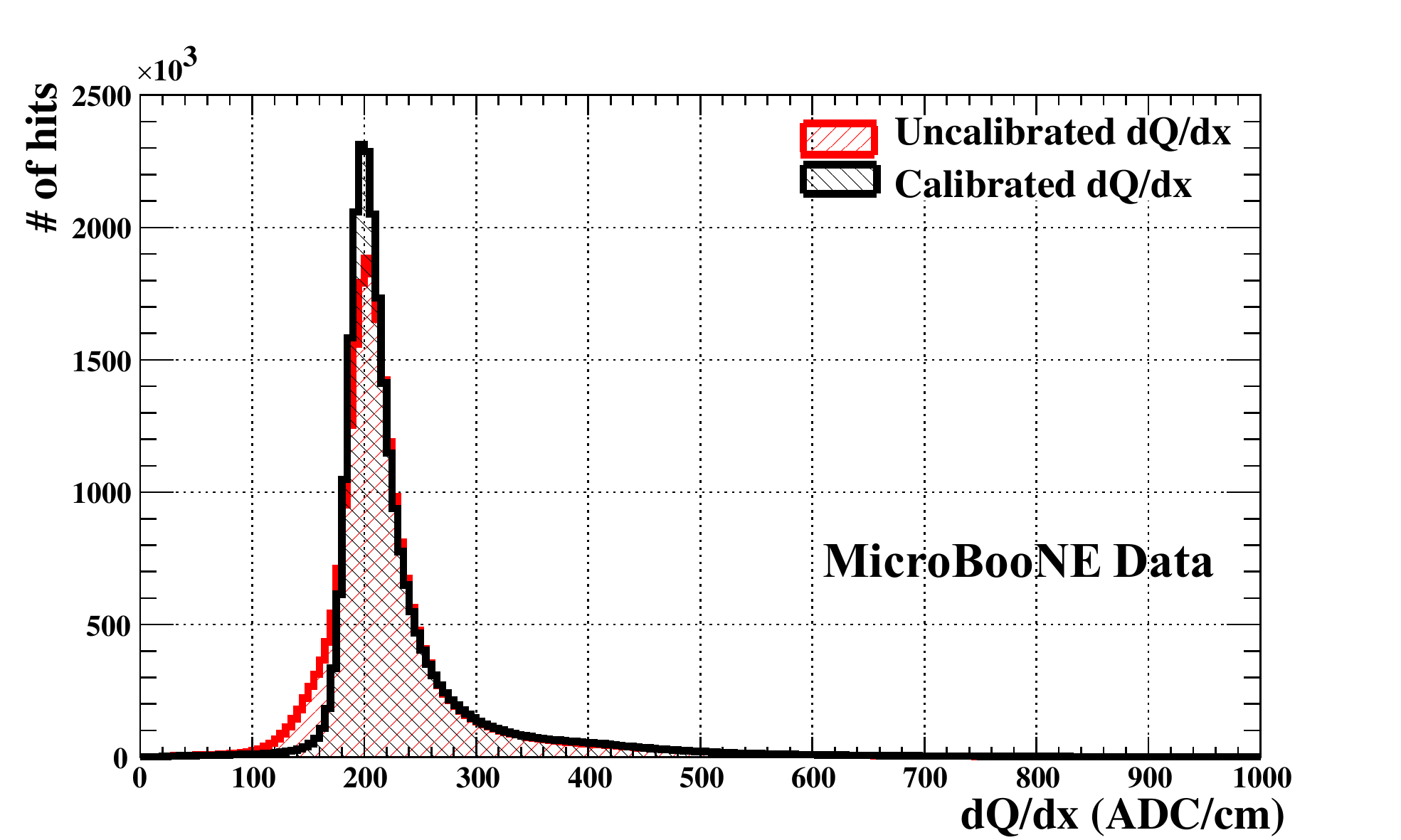}
\hspace*{0.3in}
\includegraphics[width=0.4\textwidth, height=0.265\textwidth]{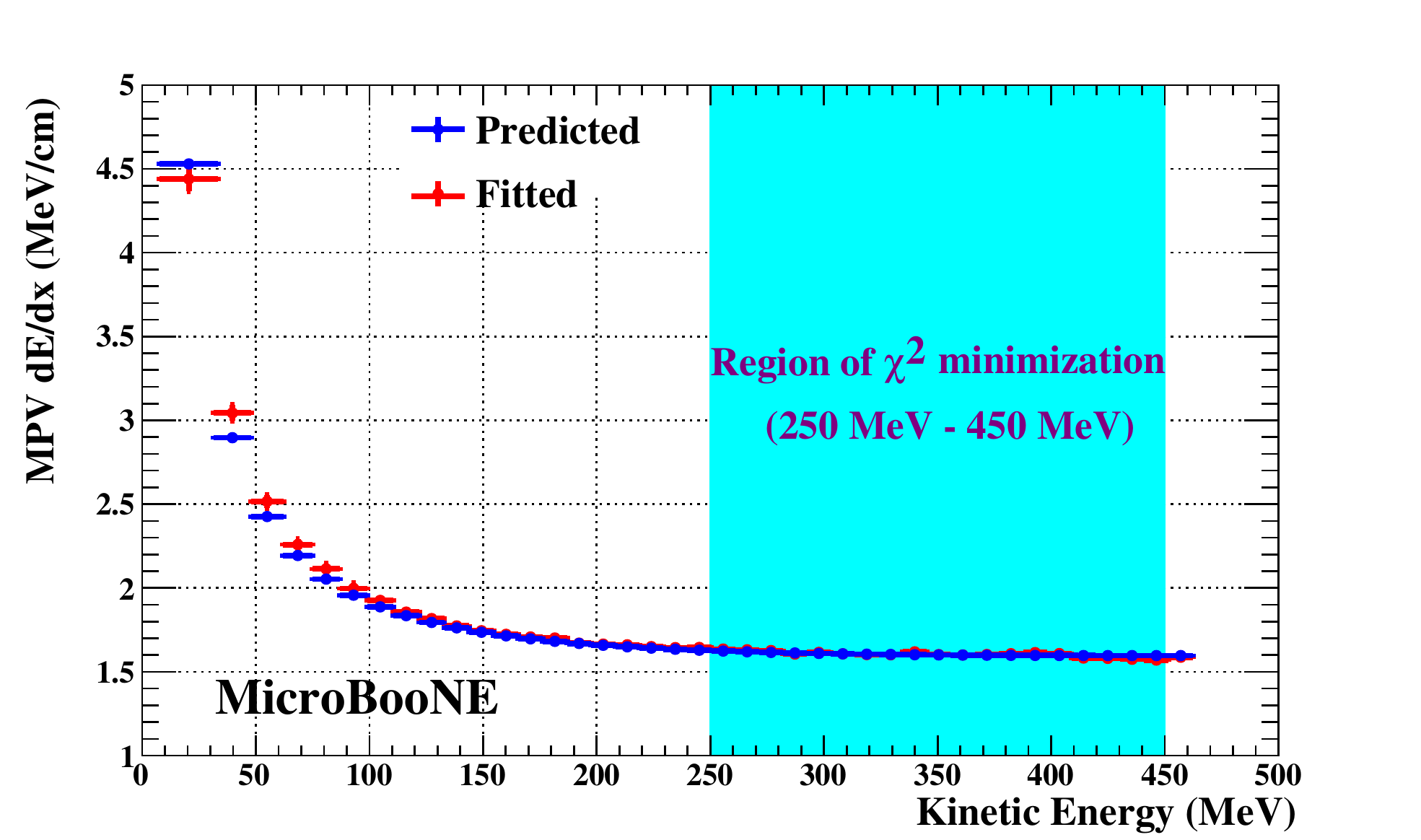}
\caption{Calibrated and uncalibrated $dQ/dx$ from crossing muons in MicroBooNE data in the collection plane (left), and comparision between the prediction and fitted most probable $dE/dx$ value for stopping muons in MicroBooNE data using the collection plane (right). Plots are taken from Ref. \cite{calibration_microboone}.}
\label{fig:dqdx_dedx}
\end{center}
\end{figure}

To obtain precise measurements of particle tracking and energy reconstruction, the following two-step calibration is performed. First, we correct the position- and time-dependence of the detector response to ionization charge using data from anode-cathode crossing cosmic muons, which span the entire drift distance and can be used to study any effects that depend on the drift distance. This process is known as the $dQ/dx$ calibration. Second, stopping muons from neutrino interactions or cosmic rays are used to study the measured and predicted most probable $dE/dx$  (energy loss per unit length) value. We then determine the calibration constant $C$ in a unit of ADC/e, which translates the corrected $dQ/dx$ in a unit of ADC/cm to $dQ/dx$ in a unit of e/cm. This process is know as $dE/dx$ calibration \cite{calibration_microboone}. Figure \ref{fig:dqdx_dedx} shows examples of calibrated and uncalibrated $dQ/dx$ from the crossing muons and the comparison between the prediction and the fitted most probable $dE/dx$ value for stopping muons in MicroBooNE data using the collection plane. The calibration of two induction planes can be accomplished independently following the same strategy.  

Stopping protons are used to further refine the relation between the measured charge and the energy loss for highly-ionizing particles. The effective recombination has been studied by comparing the measured $dQ/dx$ vs. $dE/dx$ distribution to different recombination models \cite{calibration_microboone}.

\section{Detector-Related Uncertainties}

\begin{figure}
\begin{center}
\includegraphics[width=0.38\textwidth, height=0.26\textwidth, trim = 0.1cm 0.0cm 0.0cm 0.2cm, clip]{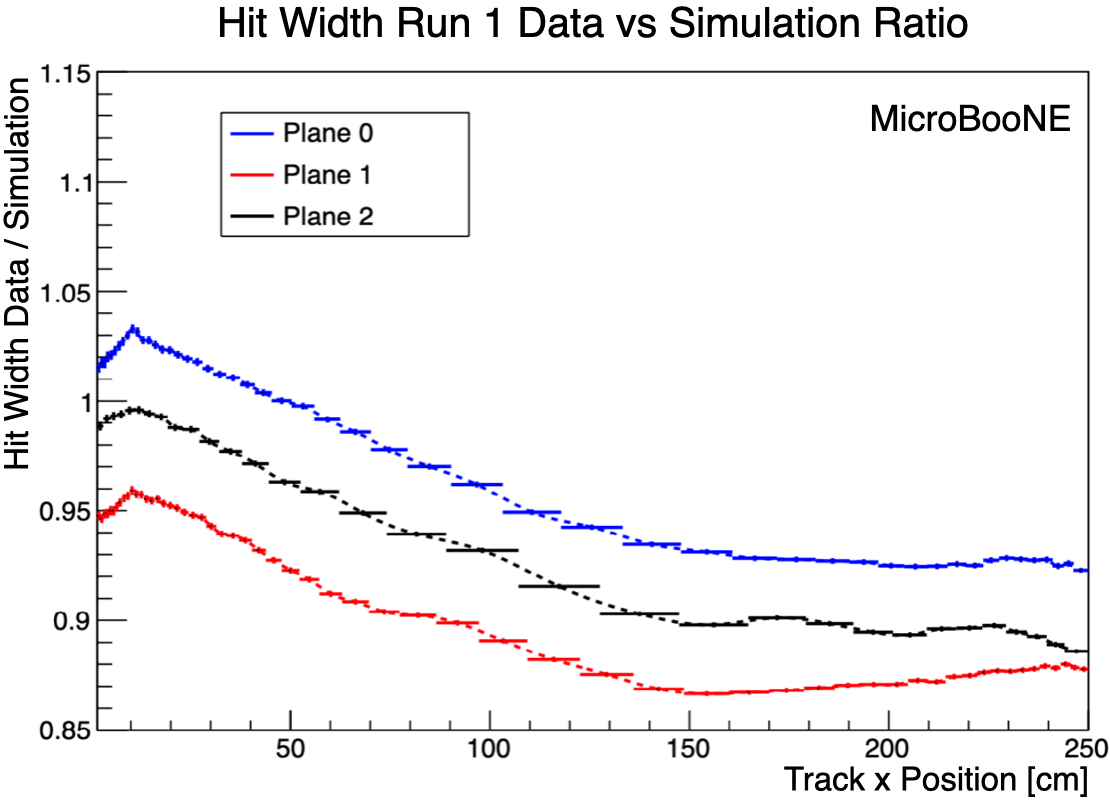}
\hspace*{0.3in}
\includegraphics[width=0.38\textwidth, height=0.275\textwidth, trim = 0.0cm 0.5cm 0.0cm 0cm, clip]{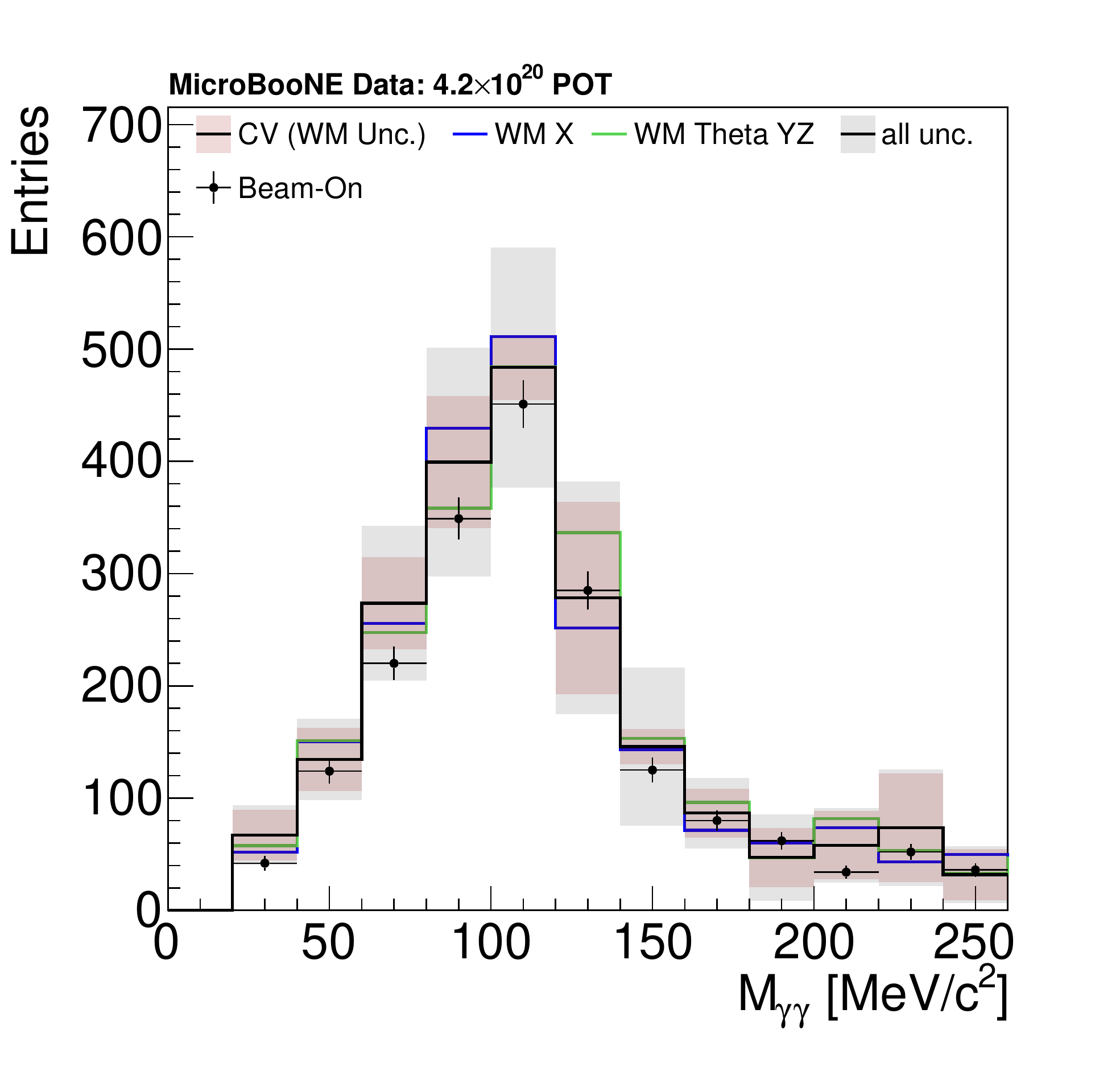}
\caption{Ratio (data/simulation) and fitted simulation modification functions for mean hit width vs. $x$ on each of the three wire planes (left), and measured diphoton invariant mass distribution using BNB beam data with uncertainties (right). Plots are taken from Ref. \cite{detector_uncertainties}.}
\label{fig:wiremod}
\end{center}
\end{figure}

To understand the detector effects and quantify the associated systematic uncertainties, modifications to simulation waveforms based on a parameterization of observed differences in ionization signals from the TPC between data and simulation are considered, in addition to other TPC variations such as electric field non-uniformities and recombination. Figure \ref{fig:wiremod} shows an example of a comparison of hit width as a function of $x$ between data and simulation, and the measured diphoton invariant mass distribution using BNB beam data with uncertainties. The novel approach for evaluating detector-related uncertainties in a LArTPC could be applied to future LArTPC detectors, such as those used in the SBN and DUNE \cite{detector_uncertainties}. 

\section{Summary}

MicroBooNE, the first operating detector in the Fermilab SBN program, has collected the world’s largest dataset of neutrino-argon interactions. The energy reconstruction and detector calibration strategies developed in MicroBooNE LArTPC can provide precision information for particle identification and measurements and are used in MicroBooNE's physics analyses such as the recent LEE results \cite{lee_microboone}. They help advance our understanding of LArTPC detector technology and have been applied to other detectors used in the SBN and DUNE.

\end{document}